\documentclass[9pt,twocolumn]{extarticle}
\pdfoutput=1

\usepackage{soul,enumerate}
\usepackage{comment}
\usepackage{titlesec} 
\usepackage[backend=bibtex,style=nature]{biblatex}
\AtEveryBibitem{\clearfield{issn}}
\AtEveryBibitem{\clearfield{isbn}}
\AtEveryBibitem{\clearfield{doi}}

\AtEveryBibitem{%
  \ifentrytype{misc}{%
    \ifboolexpr{ test {\iffieldundef{eprint}} }{%
      \clearfield{url}%
    }{}%
  }{%
    \clearfield{url}%
  }%
}
\AtEveryBibitem{\clearfield{eprint}}

\renewbibmacro*{url+urldate}{%
  \iffieldundef{url}
    {}
    {\printtext{Preprint at \printfield[noformat]{url}\setunit{\addspace}\printfield[parens]{year}}}
}

\renewbibmacro*{urldate}{}
\renewbibmacro*{date}{%
    \ifentrytype{misc}
    {}
    {\printfield{year}}%
}

\renewbibmacro*{title}{%
  \ifentrytype{misc}{%
    \printtext{\printfield[titlecase]{title}}%
  }{%
    \printtext{\printfield[titlecase]{title}}%
  }%
}

\DeclareNameAlias{sortname}{given-family}
\DeclareNameAlias{default}{given-family}
\bibliography{bib}

\usepackage{dsfont}
\usepackage{graphicx}
\usepackage{subcaption}
\usepackage[margin=0.9in]{geometry}
\usepackage[usenames,dvipsnames]{color}
\usepackage[colorlinks,linkcolor=Blue,urlcolor=Blue,citecolor=Blue]{hyperref}
\usepackage{amsmath,amssymb}
\usepackage{amsfonts}
\usepackage{bbm}
\usepackage[squaren]{SIunits}

\usepackage[table]{xcolor}
\usepackage{booktabs}
\usepackage{tabularx}
\usepackage{array}          

\usepackage{caption}
\definecolor{header}{RGB}{200,220,255}
\definecolor{row1}{RGB}{230,220,240}
\definecolor{row2}{RGB}{240,230,245}

\newcolumntype{C}{>{\raggedright\arraybackslash}X}
\newcolumntype{L}[1]{>{\raggedright\arraybackslash}p{#1}}

\usepackage{mathpazo} 
\usepackage{courier} 
\normalfont

\usepackage[T1]{fontenc}
\usepackage{caption}
\usepackage{upgreek}

\newcommand{\ket}[1]{\ensuremath{|#1\rangle\mkern-1mu}}
\newcommand{\bra}[1]{\ensuremath{\mkern-1mu\langle#1|}}

\newcommand{\dd}{\mathrm{d}}

\newcommand{\vecar}[1]{\mathop{#1}\limits^{\rightarrow}}

\newcommand{\ad}[1]{\textsuperscript{#1}\kern-2pt}

\usepackage{amsmath,amsthm,mathtools}

\renewcommand{\vec}[1]{\mathbf{#1}}

\def\({\left(}
\def\){\right)}
\def\[{\left[}
\def\]{\right]}

\makeatletter
\def\blx@maxline{77}
\makeatother

\usepackage{capt-of}

\usepackage[ruled]{algorithm2e} 
\usepackage{newfloat,algcompatible} 
\def\({\left(}
\def\){\right)}
\def\[{\left[}
\def\]{\right]}

\def\mytitle{Universal Quantum Computational Spectroscopy on a Quantum Chip  
\vspace{-3mm}}      

\title{\vspace{-1.0cm}\huge\textbf{\textrm{\mytitle}}}  
 
\author{Chonghao Zhai$^{1,\dagger}$, Jinzhao Sun$^{2,3,\dagger}$, Jieshan Huang$^{1}$, Jun Mao$^{1}$, Hongchang Bao$^{1}$, Siyuan Zhang$^{1}$,\\ Qihuang Gong$^{1,4,5,7}$, Vlatko Vedral$^{2}$, Xiao Yuan$^{6,*}$, Jianwei Wang$^{1,4,5,7*}$
}
\date{}

\begin{document}
\twocolumn[
\maketitle
\vspace{-8mm}
\begin{center}
\begin{minipage}{1\textwidth}
\begin{center}
\textit{\textrm{
\textsuperscript{1} State Key Laboratory for Mesoscopic Physics, School of Physics, Peking University, Beijing, 100871, China 
\\\textsuperscript{2} Clarendon Laboratory, University of Oxford, Parks Road, Oxford OX1 3PU, United Kingdom  
\\\textsuperscript{3} School of Physical and Chemical Sciences, Queen Mary University of London, London, E1 4NS, UK 
\\\textsuperscript{4} Frontiers Science Center for Nano-optoelectronics  \& Collaborative Innovation Center of Quantum Matter, Peking University, Beijing, 100871, China
\\\textsuperscript{5}  Collaborative Innovation Center of Extreme Optics, Shanxi University, Taiyuan, Shanxi 030006, China
\\\textsuperscript{6} Center on Frontiers of Computing Studies, School of Computer Science, Peking University, Beijing 100871, China
\\\textsuperscript{7} Hefei National Laboratory, Hefei 230088, China
\\\textsuperscript{$\dagger$} These authors contributed equally to this work. ~~~
{$\star$} emails: xiaoyuan@pku.edu.cn, jww@pku.edu.cn\\
  }}
\end{center}
\end{minipage}
\end{center}
\setlength\parindent{12pt}
\begin{quotation} 
\noindent 
Spectroscopy underpins modern scientific discovery across diverse disciplines. 
While experimental spectroscopy probes material properties through scattering or radiation measurements, computational spectroscopy combines theoretical models with experimental data to predict spectral properties, essential for advancements in physics, chemistry, and materials science.
However, quantum systems present unique challenges for computational spectroscopy due to their inherent complexity, and current quantum algorithms remain largely limited to static and closed quantum systems.
Here, we present and demonstrate a universal quantum computational spectroscopy framework that lifts these limitations. 
Through leveraging coherently controlled quantum dynamics, our method efficiently reconstructs the spectral information for both closed and open systems, furtherly for time-dependent driven systems. 
We experimentally validate this approach using a programmable silicon-photonic quantum processing chip, capable of high-fidelity time-evolution simulations. 
The versatility of our framework is demonstrated through spectroscopic computations for diverse quantum systems -- including spin systems, non-Hermitian systems, and quantum Floquet  systems -- revealing novel phenomena such as parity-time symmetry breaking and topological holonomy that are inaccessible to conventional spectroscopy or quantum eigenstate algorithms. {Furthermore, systematic benchmarking of UQCS against existing quantum algorithms is numerically performed to demonstrate its unprecedented capabilities and superior performance.} 
This work establishes a noise-robust and transformative paradigm for quantum spectral analysis.
\end{quotation}
]

\noindent 
Probing spectral characteristics of physical systems serves as a cornerstone in the fields of condensed matter physics, materials science, and biochemistry, underpinning diverse applications across modern science and technology. 
Experimental spectroscopy\supercite{118500,,Zong2023, RevModPhys.82.209} provides a standard technique for probing spectral information through external perturbations and the spectroscopic response to those excitations is analysed via spectrometers. 
For example, vibrational spectroscopy resolves molecular structures via frequency-dependent  absorption of light\supercite{118500}, and electron spectroscopy determines material electronic structures through energy-resolved measurements\supercite{Zong2023, RevModPhys.82.209}. 
%
Complementarily, computational spectroscopy provides atomic-level insights into real and hypothetical materials with precise control and inherent coherence. 
It hence enables the prediction and analysis of spectroscopic properties under experimentally challenging conditions and accelerates material discovery by pre-synthesis simulation. 
This capability holds significant importance for applications in molecular engineering\supercite{Barone2021,RevModPhys.92.015003,}, drug design\supercite{Santagati2024,}, advanced material synthesis\supercite{Louie2021}, and investigation of many-body physics \supercite{science.aag2302,Fauseweh2024}. 
%
%
Nevertheless, calculating spectroscopy of quantum systems by classical means poses significant challenges -- exponentially scaling computational resources are required for modelling quantum dynamics and in particular solving eigenstate problems. 

Quantum computers\supercite{Feynman1982} have the potential to efficiently analyze the spectroscopic  characteristics of   quantum systems 
\supercite{RevModPhys.92.015003,Santagati2024,Fauseweh2024}. 
Quantum algorithms including quantum phase estimations \supercite{Kitaev1996,science.1113479,PhysRevA.76.030306}(QPE), variational quantum eigensolvers \supercite{Peruzzo2014,science.adg9774,cerezo_variational_2021}(VQE), and their combinations \supercite{Santagati} have been proposed to prepare eigenstates on quantum computers. 
Phase estimations 
{and its generalisation in the framework of quantum signal processing\supercite{PRXQuantum.2.040203,dong2022ground}} require deep quantum circuits for enough precision to estimate eigenenergies, 
while variational methods suffer from prolonged and often inefficient optimizations  \supercite{wang_noise-induced_2021}. 
Recent advances in linear response theory\supercite{kokcu_linear_2024},  spectral filtering\supercite{Sun2025}, and algorithmic shadow spectroscopy\supercite{chan2025algorithmic}  offer alternative pathways for computing spectroscopic properties. 
However, these quantum algorithms, along with conventional experimental and computational spectroscopy,  remain constrained by their dependence on time-correlation of Hermitian operators from specific perturbations or probes. 
Consequently, 
they can only resolve excitation energies between eigenstates rather than the eigenenergies and properties of eigenstates~\supercite{kokcu_linear_2024,Sun2025,maskara2025programmable,chan2025algorithmic}. 
Furthermore, the challenge becomes significantly more pronounced for general quantum systems described by non-Hermitian Hamiltonians or time-dependent Hamiltonians. In these cases, eigenstates (as well as Fourier sideband components in  Floquet quantum systems) may lose orthogonality\supercite{ashida_non-hermitian_2020, rudner_band_2020} and novel topological phenomena could emerge in the spectral properties. Respectively, existing eigenstate characterisation falls out of the scope\supercite{ding2024,wang2023quantum,zhang2022computing}.
Overcoming these limitations represents a critical frontier for quantum computational spectroscopy, with implications for studying open quantum systems, non-equilibrium dynamics, and topological phenomena.

\begin{figure*}[ht!]
\centering 
\includegraphics[width=0.95\textwidth]{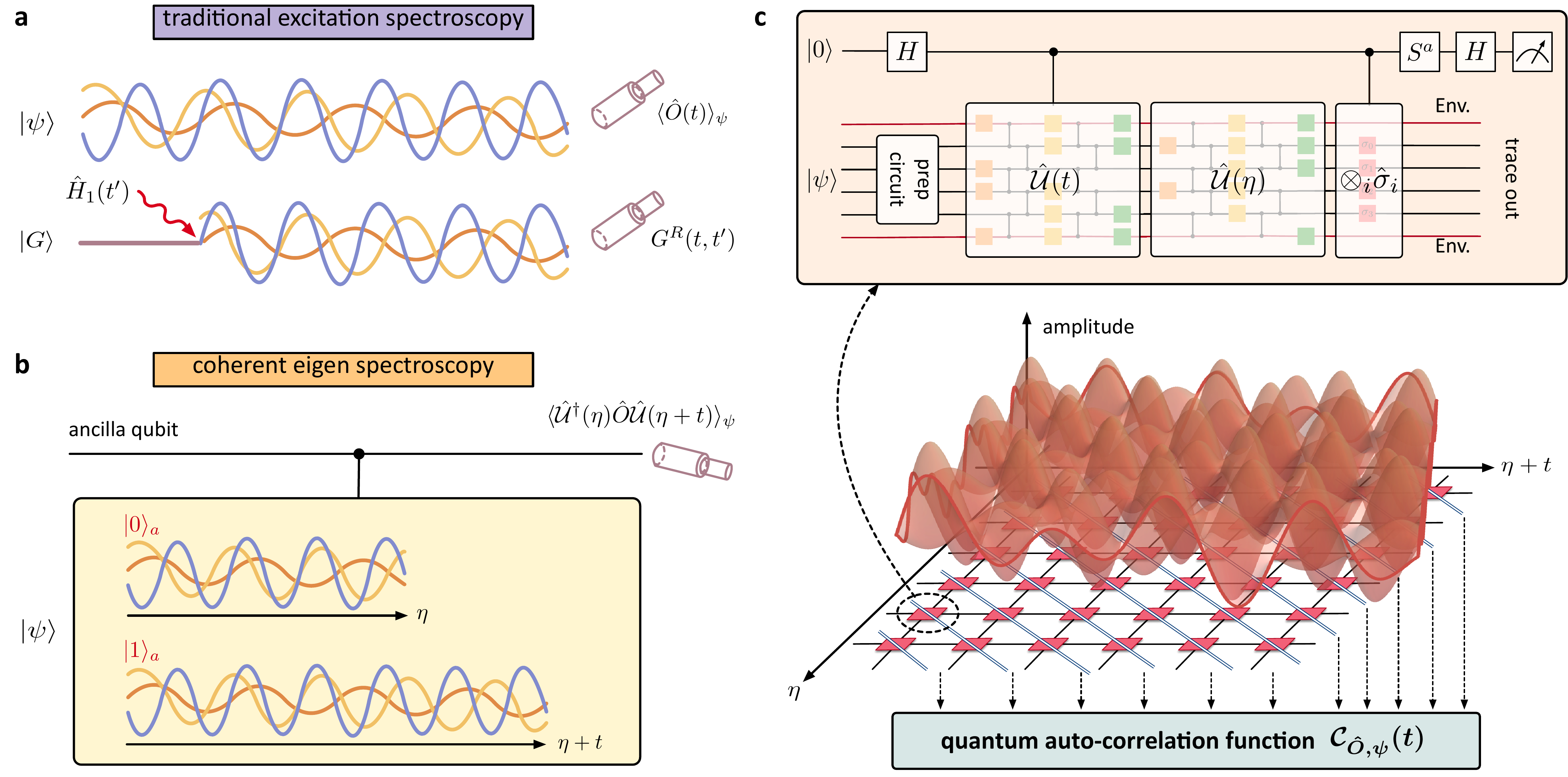}
\caption{\textbf{The framework of universal quantum computational spectroscopy.} 
Comparison of \textbf{a,} traditional excitation spectroscopy, and \textbf{b,} coherent eigen spectroscopy: In (a), traditional spectroscopy can be regarded as  excitation spectroscopy, which probes only the oscillation frequencies corresponding to eigenenergy differences. 
It relies on Fourier transforms of either probe-observable's dynamics for an initial state $\langle\hat{O}(t)\rangle_\psi=\langle\psi|\hat{U}^\dagger(t)\hat{O}\hat{U}(t)|\psi\rangle$, or, retarded Green's function $\hat{H}_1(t')$, $G^R(t,t')=-i\theta(t-t')\langle[\hat{O}(t),\hat{H}_1(t')]\rangle$ that encode linear response to perturbation $\hat{H}_1(t')$,  
where $\theta(t)$ is the Heaviside function and $[\cdot,\cdot]$ denotes the operator commutator. 
The averages in retarded Green's function are evaluated either in the ground state $\ket{G}$, or the equilibrium density matrix $\hat{\rho}=e^{-\beta\hat{H}}/Z$ with  a partition function $Z=Tr(e^{-\beta\hat{H}})$ and $\beta=1/k_BT$, where $k_B$ denotes  the Boltzmann constant and  $T$ refers to temperature. 
In (b), coherent eigen spectroscopy 
employs quantum control circuits to simultaneously evolve target dynamics through time interval $\eta$ and $\eta+t$. 
Unlike the traditional  spectroscopy, eigen spectroscopy maintains entire quantum coherence  without external excitation. 
Measurement on the ancilla qubits allows efficient estimation of the integrand of the quantum auto-correlation function $\mathcal{C}_{\hat{O},\psi}(t)$ as defined in Eq.\ref{eqn:1}. 
\textbf{c,} Quantum circuit of generalised Hadamard test for implementing ({b}).  
It consists of preparation circuits, a controlled time evolution $\hat{U}(t)$, a normal time evolution $\hat{U}(\eta)$ and a controlled Pauli-string observable $\otimes_i\hat{\sigma}_i$, $\hat{\sigma}_i\in\{\hat{I},\hat{\sigma}^x,\hat{\sigma}^y,\hat{\sigma}^z\}$. 
The time series for eigen spectroscopy calculation are sampled at discretised time stamps in the two-dimensional space of  $\eta$ and $\eta+t$, with the sum of terms along different diagonals (white lines) yielding the quantum auto-correlation function. 
The Hadamard gate is denoted as $H$ and a phase-gate $S=diag\{1,-i\}$ is applied to estimate the real ($a=0$) and imaginary ($a=1$) part of the non-Hermitian observable. 
The environment qubits (represented by red lines in circuits) are used for open quantum system simulation. 
}
\label{fig:framework}
\end{figure*}

In this work, we present universal quantum computational spectroscopy (UQCS), a framework that harnesses coherent quantum control to probe the eigenstate-resolved spectral properties of general quantum systems. Unlike conventional spectroscopic techniques, which often face challenges in resolving complex quantum systems, UQCS provides  an efficient alternative. 
The key innovation  stems from a fundamental property: the time evolution of gapped eigenstates naturally forms an orthogonal set of phase-dependent functions, whether dynamical phase or topological geometry phase, analogous to Fourier basis functions. This intrinsic phase orthogonality enables two crucial advances: (1) the cancellation of oscillating signals generated by excitations between gapped eigenstates via time-auto-correlation, and (2) the reconstruction of orthogonality to distinguish non-orthogonal gapped eigenstates and topological holonomy. 
We experimentally realise UQCS on a programmable silicon-photonic quantum chip, which can implement arbitrary controlled-unitaries with high fidelity and exceptional programmability. This device unlocks direct spectral characterisation of diverse quantum systems -- including Heisenberg spin model, non-Hermitian Hamiltonian, and periodically driven system.  We experimentally apply UQCS to observe intriguing properties. These include  gapped eigenstates reconstruction (resolving non-orthogonal eigenstates), non-Hermitian parity-time (PT) symmetry breaking (detected through spectral transitions), and topological holonomy 
in periodically driven systems (inferred from their spectral structures). Notably, UQCS achieves what was previously inaccessible using conventional spectroscopy or quantum eigenstate algorithms: it estimates arbitrary observable expectation values under gapped eigenstates in diverse quantum dynamics and deciphers intricate eigen-properties from a spectroscopic approach. 
{We perform systematic benchmarking of UQCS against existing quantum algorithms, demonstrating unprecedented capabilities and  superior performance}. 
Our method  opens new pathways to probe quantum many-body systems, open quantum systems, and non-equilibrium systems, beyond the limitations of classical techniques and existing quantum algorithms.

\vspace{2mm}
\noindent\textbf{The UQCS framework.} 
Traditional spectroscopy mainly measures excitations by probing the response dynamics of an perturbation $\hat{H}_1(t')$, characterised by either the time-varying observable  $\langle\hat{O}(t)\rangle_\psi$ 
or the retarded Green’s function $G^R(t,t')$ (Fig.~\ref{fig:framework}a). 
However, this approach 
struggles to resolve fine eigenstate properties in generic quantum systems, particularly  many-body systems, non-Hermitian systems, and non-equilibrium systems. 
To overcome these limitations, we introduce a framework based on a quantum auto-correlation function: 
\begin{equation}\label{eqn:1}
    \mathcal{C}_{\hat{O},\psi}(t) = \int \dd\eta \bra{\psi} \hat{U}^\dagger (\eta) \hat{O} \hat{U}(\eta+t) \ket{\psi}, 
\end{equation}
where $\hat{U}(t) = \mathcal{T} e^{-i\int_0^t \hat{H}(t') \dd t'}$ is the time-evolution operator for a general Hamiltonian  $\hat{H}(t)$ ($\mathcal{T}$ is the time-ordering operator), $\hat{O} $ represents a probe operator, and \ket{\psi} is the initial state ($\hbar=1$). 
The integrand in Eq.~(\ref{eqn:1}) forms a non-Hermitian time-correlation observables, which can be efficiently estimated using controlled time-evolution circuits enhanced by coherent quantum controls (Fig.~\ref{fig:framework}b). 
Quantum processor implementing the generalised Hadamard test (Fig.~\ref{fig:framework}c) enables efficient computation of quantum auto-correlation functions. The quantum circuit consists of controlled time evolution $\hat{U}(t)$, time evolution $\hat{U}(\eta)$, and 
controlled Pauli-string operators that contribute to the calculation of arbitrary observables $\hat{O}$ via Pauli decomposition. 
Finally, by applying the Fourier transform to the   $\mathcal{C}_{\hat{O},\psi}(t)$, we can resolve the frequency ($\omega$)-domain distribution of eigen-spectrum $\widetilde{\mathcal{C}}_{\hat{O},\psi}(\omega)$. 

Generally, the time evolution of any quantum state can be described as a superposition of its Fourier components: $\hat{U}(t)\ket{\psi} = \sum_n c_n e^{-iE_n t} \ket{u_n}$, where $E_n$ and $\ket{u_n}$ respectively denote the quasi-eigenenergy and quasi-eigenstate  obtained by Fourier transform, with $c_n$ as normalisation coefficients. The physical interpretation of quasi-eigenenergies and quasi-eigenstates is detailed in Methods. 
Crucially, for systems described by non-Hermitian or time-dependent Hamiltonians, these quasi-eigenstates generally form one non-orthogonal set. 
For clarity, we here focus on systems with gapped and real quasi-eigenenergies. The quantum auto-correlation function can be simplified significantly, as cross-terms between different Fourier components vanish due to their orthogonality. This results in only oscillations at the quasi-eigenenergies: 
\begin{equation}\label{eqn:2}
    \mathcal{C}_{\hat{O},\psi}(t) = 2\pi\sum_{n,m} |c_n|^2 e^{-iE_nt} \delta(E_n-E_m)\bra{u_n} \hat{O} \ket{u_n}, 
\end{equation} 
where the Dirac-$\delta$ function enforces the orthogonal formalism of phase factors. 
The quasi-eigenenergies appear as distinct peaks in the Fourier transform of the identity-operator correlation function, $\widetilde{\mathcal{C}}_{\hat{I},\psi}(\omega)$, allowing  direct extraction of any observable's expectation value via $\langle\hat{O}\rangle_{u_n}=\bra{u_n} \hat{O} \ket{u_n}/\bra{u_n}u_n\rangle = \widetilde{\mathcal{C}}_{\hat{O},\psi}(E_n)/\widetilde{\mathcal{C}}_{\hat{I},\psi}(E_n)$, where $\hat{I}$ denotes the identity operator. 

The practical implementation of UCQS requires appropriate selections of window function and time discretisation. 
To address the challenges of Dirac-$\delta$ singularities and infinite integration times in Fourier transform, we need to employ an  window function in $\boldsymbol{L}^2$-space with unit norm to extract the spectrum faithfully. We adopt the Gaussian window 
$\mathcal{G}(x,\tau) = {\exp[-x^2/(2\tau^2)]}/{\sqrt{2\pi} \tau}, \tau > 0$ in the calculation of quantum auto-correlation function ($x=\eta$) and Fourier transform ($x=t$).
This window function corresponds to an exponential approximation to the ideal Dirac-$\delta$  orthogonality in the quantum auto-correlation function: $2\pi\delta(E_n - E_m) \rightarrow \exp(-\tau^2(E_n - E_m)^2/2)$.
For reliable spectral computation, the window width $\tau$ must satisfy   $\tau \Delta E_{\min} \gg 1$, where $\Delta E_{\min}$  is the smallest  energy gap of interest.  
By truncating the time integral to the interval $[-4\tau,4\tau]$, we achieve a truncation error below 0.01\%.
Moreover, we select the time discretisation in accordance with the Nyquist-Shannon sampling theorem\supercite{shannon_communication_1949}, enabling the faithful discrete-time window Fourier transform of the time series data, where each dataset is acquired via the generalised Hadamard test experiment shown in Fig.~\ref{fig:framework}c. 
Regarding noise suppression, crucially, any oscillating noise at frequency $f_{\text{noise}}$ distinct from the quasi-eigenstates is strongly suppressed by the exponential factor $\exp(-\tau^2 (E_n - f_{\text{noise}})^2 /2)$ for each quasi-eigenenergy $E_n$. Uncorrelated noise arising from experimental quantum gate implementations  can be further mitigated through singular spectrum renormalisation of the time-series data\supercite{golyandina_singular_2013}. Given a truncation error $\epsilon_1$ and a statistical measurement error $\epsilon_2$, UQCS estimates properties of gapped eigenstates with the complexity of
$\mathcal{O}\left[({\sqrt{2\ln(1/\epsilon_1)} R(\hat{H})}/ \Delta E_{\min})^2/\epsilon_2^2 \right]$,
where $R(\hat{H})$ refers to the spectral radius of the Hamiltonian, and such complexity scales polynomially with system size for local Hamiltonians (details are provided in Supplementary Information Section \ref{sec:theory}).

Notably, the quantum auto-correlation function provides an effective resolution to the inherent challenge of quasi-eigenstate non-orthogonality. The sole requirement of   phase factor orthogonality  is satisfied for systems with quasi-eigenenergies gapped in the real part. This fundamental characteristic confers exceptional generalisation capability to our UQCS approach, enabling its application across diverse quantum systems. 
For time-independent Hermitian Hamiltonians, quasi-eigenstates become orthonormal eigenstates, allowing direct extraction of  spectral properties. 
More significantly, our framework can naturally extend to non-Hermitian Hamiltonians, where physical eigenstates maintain accessibility despite their non-orthogonal nature. 
When addressing more complex quantum dynamics governed by time-dependent Hamiltonians, quasi-eigenstates represent Fourier coefficients rather than an orthonormal basis. 
We next experimentally verify  the UQCS for both PT-symmetric non-Hermitian Hamiltonians and periodically-driven Hamiltonians, successfully identifying PT symmetry-breaking transition through their eigen-spectrum\supercite{el-ganainy_non-hermitian_2018} and topological holonomy from Floquet quasi-eigen-spectrum\supercite{rudner_band_2020,}, respectively. 

\begin{figure*}[ht!]
\centering 
\includegraphics[width=1\textwidth]{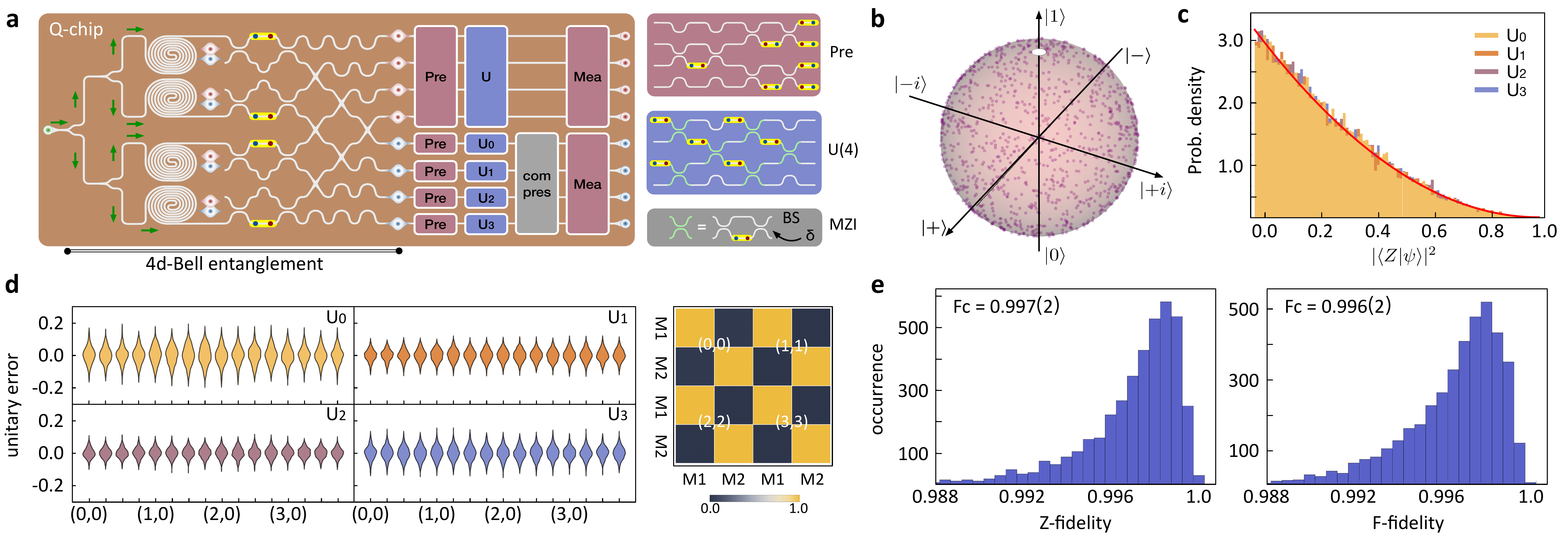}
\caption{\textbf{Characterisation of a programmable silicon-photonic quantum chip.} 
\textbf{a,} Quantum chip diagram for arbitrary controlled-unitary  U(4) operation. 
A 4-dimensional entangled Bell state of single photons is generated on-chip by simultaneously pumping four quantum light sources through the spontaneous four-wave mixing  process. The red idler photon serve as the ancilla qudit, while the blue signal photon act as the target qudit. 
Purple inset: circuit for state preparation (its reverse operation functions as a projection measurement). 
Blue inset: arbitrary unitary operations U(4) (including U$_0$-U$_3$)  are implemented using a square-arranged interferometer network of MZIs and phase shifters, which collectively realise U(2). 
Gray inset: circuit diagram of an MZI, consisting of one phase shifter and two beam-splittes (BSs) with splitting angle error $\delta$. 
The controlled-unitary operation is achieved by applying the space expansion (purple boxes) and coherent compression processes (gray box) on the target qudit state. 
\textbf{b,} Coverage of U(2) operations on the Bloch sphere. Nearly the entire sphere is accessible, except for a small forbidden region (white area) caused by beam splitter imperfections, characterized by $\delta$ with an average value of 0.019(9) for 108 MZIs. This  
leads to a forbidden area ratio $A_{FB}/{4\pi}=0.14\%$, corresponding to an  extinction ratio about 30dB. 
\textbf{c,} Experimental ($Q(x)$) and theoretical ($P(x)$) probability density of $|\bra{Z}\psi\rangle|^2$, where $|\psi\rangle$ represents 
an ensemble of four-dimensional random states  generated by Haar-random unitary sampling.  Ideally, the probability density follows $P(x)=(D-1)(1-x)^{D-2}$ (red line). The fidelity is defined as $\mathcal{F}(P,Q)=\int\sqrt{P(x)Q(x)}\dd x=0.999$ 
\textbf{d,} Measured error distributions for unitary U$_0$-U$_3$, and their measured correlation matrices. 
Each matrix element’s error is analyzed at position $(i,j)$. 
The unitaries are partitioned into two subsets (Mat1 and Mat2) for Haar-random sampling. 
\textbf{e,} Measured  classical  fidelity ($\text{F}_\text{c}$) distribution for Haar-random U(4)  operations in the computational basis (left) and the Fourier basis (right) across all four interferometer networks, showing high fidelity  operations. 
}
\label{fig:chip}
\end{figure*}

\begin{figure*}[ht!]
\centering 
\includegraphics[width=0.98\textwidth]{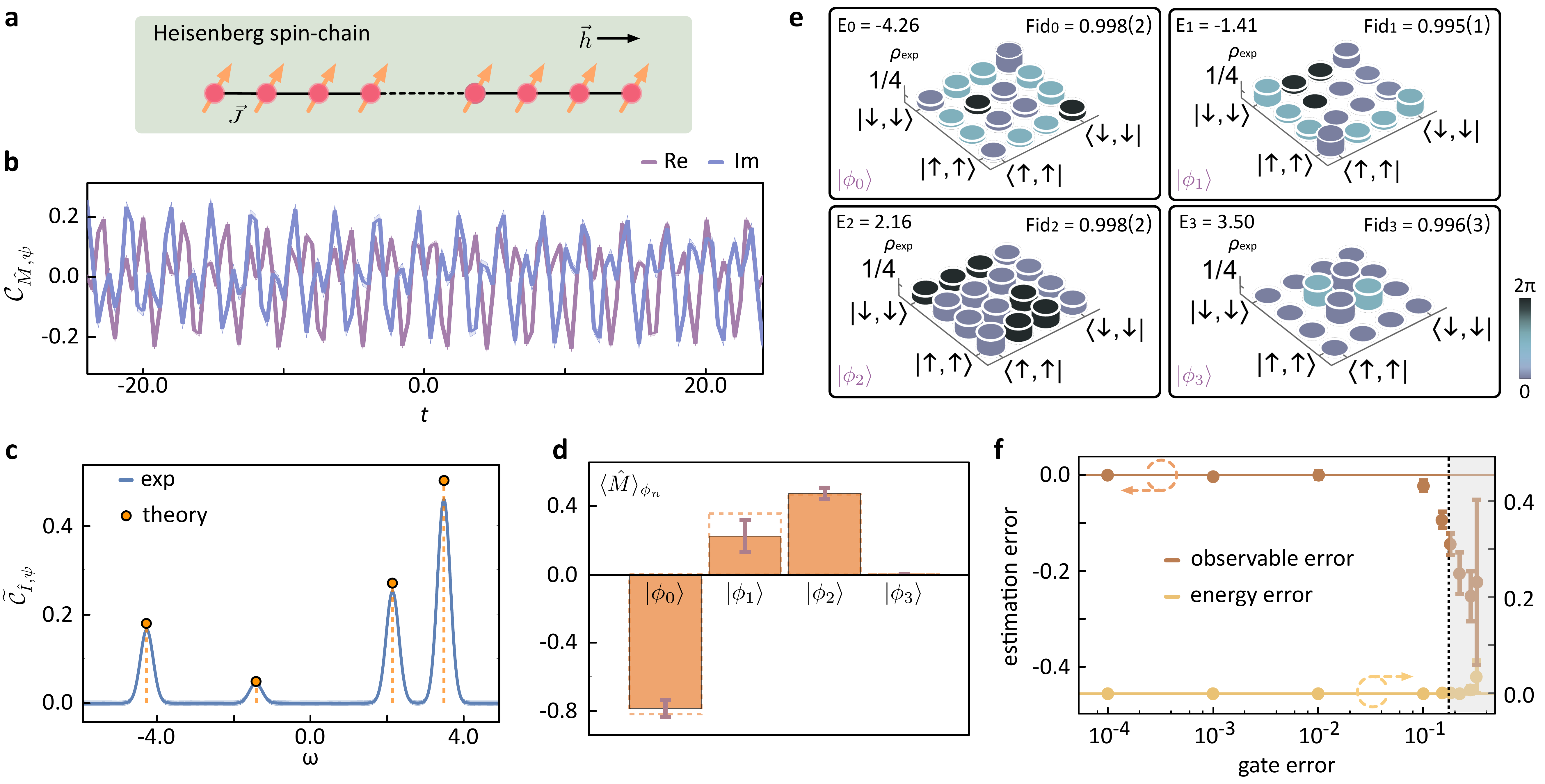}
\caption{\textbf{Benchmarking UQCS for anisotropic Heisenberg spin-chain Hamiltonians.}
\textbf{a,} Schematic of an anisotropic Heisenberg spin-chain with exchange coupling 
 $\vecar{J}$ and magnetic field $\vecar{h}$. Experimental validation adopted   a two-spin Hamiltonian. 
\textbf{b,} Measured quantum auto-correlation function of magnetisation operator $\mathcal{C}_{\hat{M},\psi}(t)$. Data points (120 time steps) are obtained through windowed auto-correlation integral of $\langle e^{i\hat{H}\eta}\hat{M}e^{-i\hat{H}(\eta+t)}\rangle_\psi$, with each point estimated from around 1000 photon-coincidence counts at discretised times $\eta$. Purple (blue) traces show the real (imaginary) parts. 
\textbf{c,} Fourier-transformed quantum auto-correlation function of identity operator $\widetilde{\mathcal{C}}_{\hat{I},\psi}(\omega) $ (blue line). 
These four Gaussian peaks represents the estimated eigenenergies centered at $\omega$ of $\{-4.257,-1.410,2.158,3.499\}$, perfectly matching with exact values $E_n=\{-4.258,-1.401,2.159,3.500\}$ (orange dashed lines). The negligible imaginary component ($\sim 3\times10^{-5}$) confirms measurement validity. 
Peak amplitudes correspond to eigenstate projection probabilities $\lvert  \langle \phi_n|\psi \rangle \rvert^2$ (theory: orange points). 
\textbf{d,} Estimation of $\langle\hat{M}\rangle_{\phi_n}$ for each eigenstate ${\phi_n}$, 
derived from the ratio of peak amplitudes of $\widetilde{\mathcal{C}}_{\hat{M},\psi}(\omega)$ and $\widetilde{\mathcal{C}}_{\hat{I},\psi}(\omega)$. Dashed boxes indicate theoretical values. 
\textbf{e,} Reconstructed density matrices for four eigenstates ${\phi_n}$ by using the UQCS-enabled quantum state tomography.  $\mathrel{|}\downarrow\rangle,\mathrel{|}\uparrow\rangle$  denote the spin-qubit basis states. 
\textbf{f,} Error analysis for an 8-site spin-chain system implemented with UQCS in noisy generalised Hadamard test circuits.  
We examine the ground state energy and spin-correlation operator $\hat{\sigma}^z_1\hat{\sigma}^z_5$. Gate errors ($\epsilon_g$) are modeled as noise following the  circularly-symmetric complex Gaussian distribution $\mathcal{C}\mathcal{N}(0,\epsilon_g/(N/2))$
applied to each element of the single-step time-evolution unitary $e^{-i\hat{H}\delta t}$ ($\delta t=8\tau/N, N=120$). The projection probability of ground state is denoted by the black dashed line. When the gate error exceeds the projection probability, the estimation error of ground state energy and spin-correlation increases fast (gray area).  All error bands in (b,c) and error bars in (d) denote the $\pm \sigma$ intervals, estimated by the Monte Carlo simulation for Poissonian sampling noise in the experiment, and by 10 independent noisy circuits in simulation in (f). 
}
\label{fig:hermitian}
\end{figure*}

\begin{figure*}[t]
\centering 
\includegraphics[width=1\textwidth]{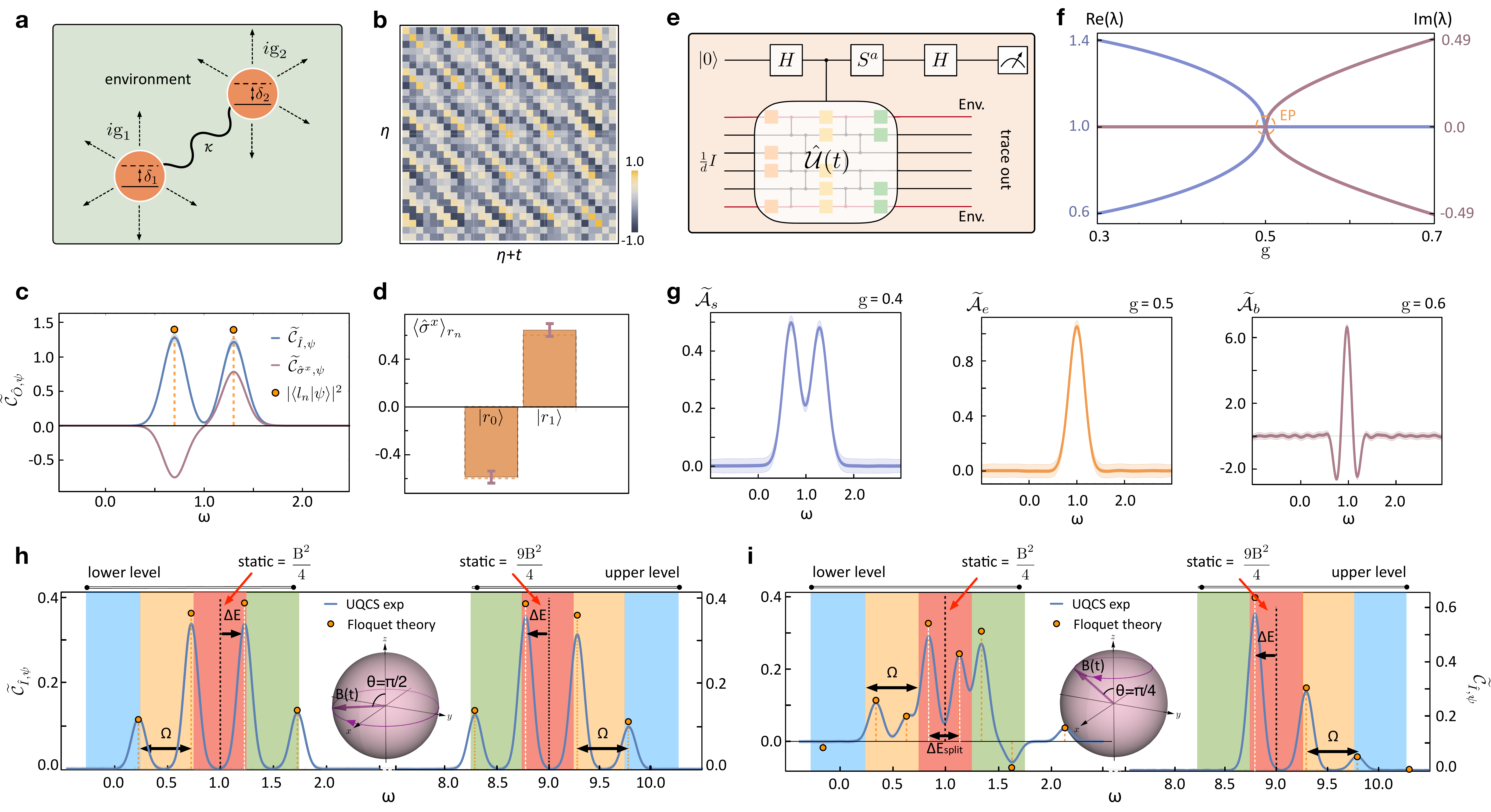}
\caption{\textbf{Benchmarking UQCS for general dynamics in non-Hermitian and Floquet quantum systems.}
\textbf{a,} 
Schematic of a non-Hermitian two-mode quantum system. 
\textbf{b,} Measured quantum auto-correlation function $\mathcal{C}_{\hat{I},\psi}(t)$   for the identity operator as $\langle e^{i\hat{H}_{NH}^\dagger\eta} e^{-i\hat{H}_{NH}(\eta+t)}\rangle_\psi$, where $g=0.4$ (partial samples are shown for clarity). 
\textbf{c,} Fourier-transformed eigen-spectra of identity operator $\widetilde{\mathcal{C}}_{\hat{I},\psi}(\omega) $ (blue lines) and Pauli-$X$ operator $\widetilde{\mathcal{C}}_{\hat{{\sigma}}^x,\psi}(\omega)$ (purple lines), peaking at $\{0.700,1.302\}$ (exact value $\lambda_n^*=\{0.7,1.3\}$, orange points), with amplitudes approximating $|\langle l_n|\psi\rangle|^2$.  
\textbf{d,} Estimation of $\langle\hat{\sigma}^x\rangle_{r_n}$ for each physical eigenstate, 
derived from the ratio of peak amplitudes in (c). Dashed boxes indicate exact theoretical values. 
\textbf{e,} The variant UQCS circuits to track the PT phase transition. The utilizing of maximally mixed initial state ($\rho = I/d$) overcomes the plight of non-diagonalisability. 
\textbf{f,} The change of eigenenergies in PT phase transition.  The real (blue line) and imaginary (purple line) parts of the eigenenergies coalesce at the EP. 
\textbf{g,} Fourier-transformed eigen-spectra $\widetilde{\mathcal{A}}_s$ (before EP, g=0.4), $\widetilde{\mathcal{A}}_e$ (at EP, g=0.5), $\widetilde{\mathcal{A}}_b$ (after EP, g=0.6) of the samples from variant UQCS circuits in (e).
\textbf{h} and \textbf{i}, Computation of Floquet spectrum and topological holonomy of periodically-driven  systems. 
Fourier-transformed eigen-spectra of the identity operator for a nuclear quadrupole resonance Hamiltonian with a periodically driven magnetic field  $\vec{B}(t)$ at angular frequency  $\Omega$: ({h}), magnetic field circulates along the equator ($\theta=\pi/2$), and  ({i}),  magnetic field follows a polar circle ($\theta=\pi/4$), where $\theta$ denotes the zenith angle. Blue lines show experimental UQCS results; orange points denote Floquet theoretical predictions. 
The static system ($\Omega=0$) has eigenenergies $B^2/4$ and $9B^2/4$ (black dashed lines). Periodically driving generates the Floquet-Bloch bands with $\Omega$-spacing (colored regimes). 
In (h), the energy shift $\Delta E$ (black arrows) from the static eigenenergy to the first-band single quasi-eigenenergy (white lines, for both lower and upper levels) enables estimation of the U(1) holonomy (Berry phase). 
In (i), the upper level retains a single quasi-eigenenergy band structure, while the lower level exhibits two distinct quasi-eigenenergies with splitting $\Delta E_\text{split}$. This splitting signifies genuine SU(2) holonomy, allowing extraction of the gauge-invariant Wilson loop. All error bands with spectrum lines denote the $\pm \sigma$ intervals, estimated by the Monte Carlo simulation for  Poissonian sampling noises. 
}
\label{fig:general}
\end{figure*} 

\begin{figure*}[ht]
\centering 
\includegraphics[width=0.7\textwidth]{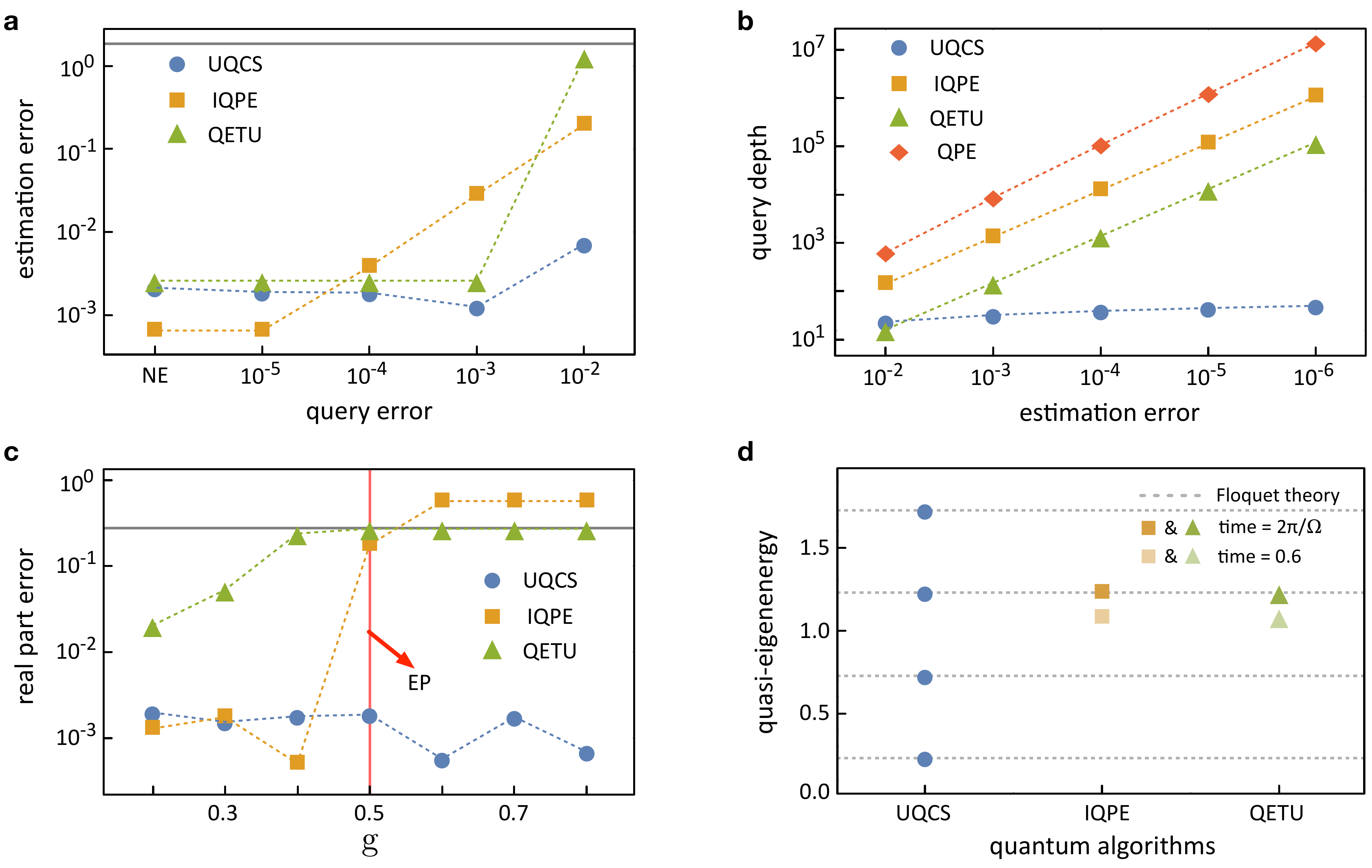}
\caption{\textbf{{Benchmarking quantum algorithms for spectroscopic estimation of general  systems.}}
{\textbf{a,} and \textbf{b,} Ground-state energy estimation for the 8-site spin chain Hamiltonian:  (a), energy estimation error ($\epsilon$)  versus query error ($\epsilon_q$), and (b), query depth versus target estimation error $\epsilon$ ($\epsilon_q$ = 0). 
For IQPE and QETU, we apply a shift of eigen-spectrum  $\hat{H}\delta t\rightarrow\hat{H}\delta t+11*\hat{I}$, to ensure the rescaled ground-state energy falls within $(0,2\pi)$. 
To achieve comparable estimation accuracy when no query error (NE), the query depths are chosen to \{$30,~2^{11},~ 274$\} for UQCS, IQPE and QETU, respectively. 
\textbf{c,} Estimation of eigenenergies (real part) in non-Hermitian system. 
While UQCS can estimate the non-Hermitian eigenenergies across the PT-symmetry-breaking transition, 
IQPE fails when passing through the EP at $g=0.5$, and QETU remains invalid throughout the entire parameter regime. 
\textbf{d,} Estimation of  quasi-eigenenergy in Floquet quantum system.  The  dashed horizontal lines denote 
 theoretical quasi-eigenenergies of the Floquet  system. 
 While UQCS can resolve all oscillatory components associated with distinct quasi-eigenenergies, both IQPE and QETU are constrained to accurately estimating only a single quasi-eigenenergy, provided the evolution time of the query unitary matches the driving period  ($2\pi/\Omega$).  
IQPE: iterative quantum phase estimation; QETU: quantum eigenvalue transformation of unitary matrices. In (a) and (c), The gray line indicates the initial guess bound for ground-state energy or the real part of eigenenergy. 
In (a)-(d), 
the initial state has $\zeta =0.9$ overlap with the target eigenstate. Simulation results are obtained using MindSpore Quantum with 1000 shots per run.}}
\label{fig:benchmark}
\end{figure*} 

\begin{table*}[ht]
\caption{\textbf{{Comparison of quantum algorithms for spectroscopic estimation of various quantum systems.}}}
\label{tab:benchmark}

\vspace{-2mm}

\begin{center}
\renewcommand{\arraystretch}{1.2} 
\begin{tabularx}{\textwidth}{L{4.5cm}L{4.5cm}L{4cm}CC}
  \toprule
  \rowcolor{header} 
  \textbf{Quantum algorithms} & \textbf{Quantum systems} & \textbf{Query depth\textsuperscript{a}}  & \textbf{Excitation- or eigen-energy} & \textbf{State properties} \\
  \midrule
  \rowcolor{row1}
  QPE\supercite{Kitaev1996,science.1113479} & Hermitian TI & $\mathcal{O}(\epsilon^{-1}\log(\epsilon^{-1}))$ & eigen & Yes  \\
  \rowcolor{row2}
  IQPE\supercite{PhysRevA.76.030306} & Hermitian TI, PT exact phase & $\mathcal{O}(\epsilon^{-1})$ & eigen & No   \\
  \rowcolor{row1}
  QETU (prepare ground state)\supercite{dong2022ground} & Hermitian TI & $\mathcal{O}(\zeta^{-2}\Delta^{-1}\log(\epsilon^{-1}\zeta^{-1}))$ & eigen & Yes \\
  \rowcolor{row2}
  QETU (estimate eigenenergy)\supercite{dong2022ground} & Hermitian TI & $\mathcal{O}(\epsilon^{-1}\log(\zeta^{-1}))$ & eigen & Yes \\
  \rowcolor{row1}
other QCS\supercite{Sun2025,chan2025algorithmic} & Hermitian TI & $\mathcal{O}(\Delta^{-1}\sqrt{\log(\epsilon^{-1})})$ & excitation & No  \\
  \rowcolor{row2}
VQE\supercite{Peruzzo2014,science.adg9774,cerezo_variational_2021} & Hermitian TI\textsuperscript{\textbf{b}} &  ansatz-dependant  & eigen & Yes \\
    \rowcolor{row1}
  UQCS (this work) & Hermitian TI, non-Hermitian, Floquet (general dynamics) & $\mathcal{O}(\Delta^{-1}\sqrt{\log(\epsilon^{-1}\zeta^{-1}\Delta)})$ & eigen & Yes \\ 
  \bottomrule
\end{tabularx}
\end{center}
\vspace{-1mm}
{\small{IQPE: iterative quantum phase estimation; QETU: quantum eigenvalue transformation of unitary matrices; VQE: variational quantum eigensolver; EP: exceptional point;  TI: time-independent quantum system;  
\textbf{a}: query depth accounts only for the parameters discussed in the corresponding references; 
\textbf{b}: it remains unknown  whether VQE can solve non-Hermitian eigenproblems.}}
\end{table*}

\vspace{2mm}
\noindent 
\textbf{A programmable silicon-photonic quantum chip.} While the UQCS is platform-independent and can be implemented on any quantum computing platform, we  here validate   it on a programmable silicon-photonic quantum processor capable of  implementing controlled dynamics with high-degree arbitrariness and high-level fidelity. The quantum chip illustrated   in Fig.~\ref{fig:chip}a is devised to implement the generalised Hadamard test in Fig.~\ref{fig:framework}c, the core operation of UQCS, and fabricated by complementary metal oxide semiconductor (CMOS) processes (see Methods). 
Harnessing multidimensional quantum operations of integrated quantum photonic technologies\supercite{Zheng,Chi2022}, this  chip is able to generate  four-dimensional Bell state of signal and idler photons as $\ket{\Phi}_\text{Bell}$= $\frac{1}{\sqrt{d}} \sum_{i=0}^{d-1} \ket{i} \ket{i}$,  $d$= 4. Then applying the Hilbert space expansion and coherent compression processes on the entangled state, we obtain the controlled-unitary operation as $\frac{1}{\sqrt{d}}\sum_{i=0}^{d-1}\ket{i}_a \text{U}_i \ket{\psi}_T$, where 'a' denotes the ancilla qudit and 'T' denotes the target qudit $\ket{\psi}_T$. Eventually, the chip   implements a multi-valued controlled-unitary quantum circuit, where four four-dimensional unitary U(4) can be arbitrarily  operated. In experiment, we set $\text{U}_0,\text{U}_1$ as $\hat{U}(\eta)$ and $\text{U}_3,\text{U}_4$ as $\otimes_i\hat{\sigma}_i\cdot\hat{U}(\eta+t)$. 
This duplicated setting of Hadamard test can effectively mitigate errors arising from device nonuniformity \supercite{marshman_passive_2018}, by the adoption of four-dimensional qudit in the ancilla register. 
We first characterised the arbitrariness and fidelity of the controlled unitary realised on the quantum chip.

Arbitrary unitary operations are implemented via a square network of integrated-optical Mach-Zehnder interferometers (MZIs)\supercite{Oxford:U}, as shown in Fig.~\ref{fig:chip}a. The beamsplitter (BS) angle error  $\delta$ in each MZI constrains the operation to a subspace of the Bloch sphere (Fig.~\ref{fig:chip}b).
To assess the interferometer network’s performance in UQCS, we evaluate several key metrics. First, we characterise the coverage $Cr$, which quantifies the likelihood of successfully implementing an arbitrary unitary matrix with the U(4) network. 
For a $d\times d$  network, it can be measured by $Cr=e^{-d^3\delta^2/3}$ under the uncorrelated Gaussian error model \supercite{hamerly_asymptotically_2022}. In our U(4) network device, it achieves a coverage of 99.2\%, demonstrating near-whole operation range. 
We next characterise the error uniformity, that evaluates whether an uncorrelated uniform error distribution model could apply across all unitary settings. 
This is quantified by the fidelity between experimental and theoretical probability distributions for projections onto the computational basis states  $|\bra{Z}\psi\rangle|^2$ ($Z$=\{0,1,2,3\}), using an ensemble of Haar-random quantum states generated by Haar-random unitary matrices\supercite{choi_preparing_2023}. 
As shown in Fig.~\ref{fig:chip}c, experimental results from four U(4) networks exhibit strong agreement with theoretical predictions across 1000 random unitaries per network, achieving a high fidelity. 
Furthermore, we analyse the error distribution and correlations across all U(4) networks sampled from the Haar-random unitary ensemble in Fig.~\ref{fig:chip}d. All matrix element errors remain bounded within $\pm0.2$ with a standard deviation of 0.06. When examining random bipartitions of the ensemble, the errors in the four diagonal matrix elements appear to be nearly uncorrelated. 
Another crucial aspect is the classical unitary operation fidelity, which is quantified through a modified Hilbert-Schmidt inner product $F_c(U_{thy},U_{exp})=\frac{1}{d}Tr(|U_{thy}^\dagger||U_{exp}|)$, where 
$|\cdot|$ denotes element-wise absolute values. 
As reported in Fig.~\ref{fig:chip}e, the fidelity distributions for both computational and Fourier bases consistently exceed 0.996. 

These results show that the superior performance of our quantum chip allows for  high-fidelity implementation of nearly arbitrary controlled time-evolution operators. Crucially, the operational noise demonstrates both uniformity and absence of correlations across different unitary operations -- characteristics that permit effective noise suppression through singular spectrum renormalisation, as previously described.

\vspace{2mm}


\noindent\textbf{Time-independent Hermitian quantum systems.} 
To validate the UQCS and assess its noise robustness, we first benchmark its performance by studying spin-chain dynamics implemented on our quantum photonic chip. As a test case, we analyse the spectral properties of an anisotropic Heisenberg spin-chain Hamiltonian subject to an external magnetic field (Fig. \ref{fig:hermitian}a): 
\begin{equation}
    \hat{H}=\sum_{\alpha,i}J_\alpha \hat{\sigma}_i^\alpha\hat{\sigma}_{i+1}^\alpha+\sum_{\alpha,i}h_\alpha\hat{\sigma}_i^{\alpha}, 
\end{equation}
where $\hat{\sigma}_i^{\alpha}$ ($\alpha=x,y,z$, $i=1,2$) is the Pauli operator on the $i$-th site, $\vecar{J}=[-1,-1,-1.5]$ denotes  the exchange coupling, and $\vecar{h}=[1.5,0,0.5]$ is the magnetic strength. 
We target at revealing the spectrum of the  z-axis  magnetisation operator $\hat{M}=\hat{\sigma}^z_1\otimes\hat{I}_2+\hat{I}_1\otimes\hat{\sigma}^z_2$, expressed in its Pauli string decomposition.  
To achieve optimal energy resolution, we opt for a broad Gaussian window function with $\tau=6$ for our spectral analysis.
The upper bound of spectral radius  follows directly from $R(\hat{H})\leq\sum_{\alpha,i}|J_\alpha|+\sum_{\alpha,i}|h_\alpha|$, which yields a corresponding lower bound for the sampling frequency $f_s\geq\frac{15}{2\pi}s^{-1}$. 
On the quantum chip, the initial state is prepared in the product state as $\ket{\psi}_T=\mathrel{|}\uparrow\rangle\otimes\mathrel{|}\downarrow\rangle$ 
and the networks are mapped accordingly for the Hadamard test. 
%
%
%
We measured the quantum auto-correlation functions  $\mathcal{C}_{\hat{M},\psi}(t)$ (see Fig.~\ref{fig:hermitian}b) and $\mathcal{C}_{\hat{I},\psi}(t)$ (see Fig.~S3), 
when choosing a time discretisation of 120 stamps for each $t$. 
Fourier transform of the experimental data results in the eigen-spectra for the identity operator $\widetilde{\mathcal{C}}_{\hat{I},\psi}(\omega) $ in Fig.~\ref{fig:hermitian}c. 
In the spectral distribution, the Gaussian peak amplitudes correspond to the measured projection probabilities of the initial state onto each eigenstate, while the peak centers determine the estimation of eigenenergies -- both showing excellent agreement with theoretical predictions. 
From the spectral ratio of $\widetilde{\mathcal{C}}_{\hat{M},\psi}(\omega)/\widetilde{\mathcal{C}}_{\hat{I},\psi}(\omega)$, the magnetization expectation values $\langle\hat{M}\rangle_{\phi_n}$ can be estimated, as reported in Fig.~\ref{fig:hermitian}d, where the evaluated error for $\langle\hat{M}\rangle_{\phi_n}$ arises as its projection probability near the gate-error floor. 
In addition, since the expectation value of any observable can be estimated using UQCS, we applied this method to perform complete quantum state tomography. This allows for the reconstruction of density matrices for all eigenstates, as demonstrated in Fig.~\ref{fig:hermitian}e. The results show high fidelities of $\text{Fid}=\{0.998(2),0.995(1),0.998(2),0.996(3)\}$, defined as $(\text{Tr}[\sqrt{\sqrt{\rho_0}\cdot \rho \cdot \sqrt{\rho_0}}])^2$, where $\rho_0$ and $\rho$ are ideal and measured eigenstates, respectively, confirming the accuracy of this approach. 

We further investigate the noise resistance of UQCS as the system size scales. For those dominant noise sources, that are uncorrelated Gaussian errors in controlled unitary operations and shot noise in measurements, UQCS shows strong robustness when the defined gate error remains below the projection probability of the target eigenstate. 
To benchmark this, we simulate noisy generalised Hadamard test circuits for an 8-site spin chain with periodic boundary conditions using MindSpore Quantum\supercite{xu2024mindspore}. The system is initialised in the trial state $\ket{\psi}_T=\mathrel{|}\downarrow\rangle^{\otimes8}$, and we estimate both the ground-state $\ket{G}$ energy and the spin-correlation $\bra{G}\hat{\sigma}^z_1\hat{\sigma}^z_5\ket{G}$ between the two farthest sites. 
We simulated noisy controlled dynamics (non-unitary with error) at varying error levels, sampling 1000 shots per generalised Hadamard test. The results are shown in Fig.~\ref{fig:hermitian}f. 
After renormalising the singular spectrum of the sampled quantum auto-correlation time series,   a small estimation error is observed until the noise approaches the ground state’s projection probability, as shown in Fig.~\ref{fig:hermitian}f. These simulation results confirm that UQCS reliably estimates eigen-spectra and eigenstate properties, provided the target eigenstate’s projection probability exceeds the gate error level. 

\vspace{2mm}
\noindent\textbf{Non-Hermitian quantum systems.} 
For open quantum systems, the interaction with the environment could induce the non-Hermitian effective Hamiltonians $\hat{H}_{NH}$, which cannot be diagonalised by unitary like closed quantum systems. That forces a general diagonalisation in dual linear space, $\hat{H}_{NH}=\sum_i\lambda_i\ket{r_i}\bra{l_i}$ with $\lambda_i \in \mathbb{C}$, non-orthogonal right eigenvectors $\ket{r_i}$ (physical states) and their dual left counterparts $\bra{l_i}$\supercite{ashida_non-hermitian_2020}.  
A notable exception is parity-time (PT) symmetric systems, which have attracted significant attention for their ability to retain real eigenenergies when in `PT exact phase'\supercite{PhysRevLett.80.5243,el-ganainy_non-hermitian_2018,Feng2017, Özdemir2019}. It provides an ideal testbed for UQCS to probe open quantum system dynamics with still existing phase orthogonalities. 


To illustrate the general capability of  UQCS, we experimentally implement a prototypical non-Hermitian system comprising two coupled quantum modes with environmental interactions (Fig.~\ref{fig:general}a): 
\begin{equation}
    \hat{H}_{NH}=\begin{pmatrix}
    \delta_1-ig_1 & \kappa \\
    \kappa & \delta_2+ig_2
    \end{pmatrix}
\end{equation}
where $\delta_1,\delta_2$ are the detuning of two quantum modes, $g_1,g_2$ represent gain or loss for each quantum mode and $\kappa$ is the coupling coefficient in between. 
In our case, we chose $\delta_1=\delta_2=1.0$, $g_1=g_2=g$ and $\kappa=0.5$. 
When $g<\kappa$ (take $g=0.4$), the eigenenergies are real. 
We encode this two-mode non-Hermitian system in two optical modes ($\ket{1}_{d=4}$ and $\ket{2}_{d=4}$) in the chip's 4-dimensional Hilbert space. The non-unitary time evolution $e^{-i\hat{H}_{NH}t}$ is embedded within U(4) operations, initialised with state  $\ket{\psi}_T=\ket{1}_{d=4}$. 
For this system, we select $\tau=6$  and sample the quantum auto-correlation function at 30 time steps for integration and Fourier analysis. 
In experiment, the time correlations $\langle e^{i\hat{H}_{NH}^\dagger \eta}e^{-i\hat{H}_{NH} (\eta+t)}\rangle_\psi$ are measured on the quantum chip, the real part of which are partially reported in the matrix in Fig.~\ref{fig:general}b. 
The calculation of quantum auto-correlation function $\mathcal{C}_{\hat{I},\psi}(t)$ is realised by the windowed sum along different matrix diagonals. In addition, we also measure correlations for the parity-like operator $\hat{\sigma}^x$ in 'PT exact phase', which exhibits a pseudo-Hermitian relationship $\hat{\sigma}^x\hat{H}_{NH}\hat{\sigma}^x=\hat{H}_{NH}^\dagger$ and flips sign under physical eigenstates $\ket{r_n}$. 
Fourier transform of these correlations reveals spectral peaks centered at the system’s real eigenenergies (Fig.~\ref{fig:general}c). Crucially, the peak amplitudes in $\widetilde{\mathcal{C}}_{\hat{I},\psi} (\omega)$ correspond to $|\bra{l_n}\psi\rangle|^2$ rather than conventional projection probabilities, reflecting the non-Hermitian nature of the system. Similar to the previous case, calculating the ratio of peaks between the two spectra returns the expectation values of $\hat{\sigma}^x$  (Fig.~\ref{fig:general}d),  in good agreement with theoretical predictions.

Figure~\ref{fig:general}e illustrates a variant UQCS quantum circuits with a single controlled time evolution and the maximally mixed initial state ($\rho = I/d$) to reveal the PT phase transition, where the system's eigenenergies change from purely real to complex-conjugate pairs (shown in Fig.~\ref{fig:general}f). At the exceptional point ($g=\kappa$), $\hat{H}_{NH}$ becomes non-diagonalisable  and admits only a Jordan canonical form, that is the hallmark of critical point in PT phase transition. 
The variant UQCS circuits can still extract the diagonal terms' oscillations even if there only exists a reduced Jordan canonical form for Hamiltonians, which represents challenging for conventional techniques.
By measuring the ancilla qudit at 30 discrete time stamps, we obtain the time series $\mathcal{A}(t)=Tr(e^{-i\hat{H}_{NH}t})/d$ for different $g$ values, which captures oscillations associated with eigenenergies along the principal diagonal of the Jordan matrix. 
The Fourier transformed spectra are obtained and reported in Fig.~\ref{fig:general}g, which displays three distinct regimes: in PT exact phase ($g=0.4$, $\widetilde{\mathcal{A}}_s$), at exceptional point ($g=0.5$, $\widetilde{\mathcal{A}}_e$) and in PT broken phase ($g=0.6$, $\widetilde{\mathcal{A}}_b$). 
The Gaussian peaks in all spectra are centered at the real parts of eigenenergies, with UQCS clearly capturing their coalescence at the exceptional point. Note that the negative amplitudes and oscillating behavior in the PT-broken phase spectrum directly result from non-vanishing imaginary part of eigenenergies, serving as clear signatures of phase transition.

\vspace{2mm}
\noindent\textbf{Periodically driven quantum systems.}
%
The drive-induced non-equilibrium dynamics would bring out new phenomena not captured by the original system, such as Floquet topogical phases and Floquet topological insulator with quasi-eigenenergies band structures\supercite{zhang_digital_2022,rudner_band_2020}.
Conventional approaches for static eigen-problems fail to resolve these emergent topological features. 
UQCS overcomes this limitation by directly probing spectral signatures in quantum dynamics. Its phase-orthogonality principle naturally accommodates time-dependent Hamiltonians $\hat{H}(t)$: driving fields in Hamiltonians  can be regarded as phase modulation of quantum states, enabling efficient extraction of oscillating components. 

We investigate nuclear quadrupole resonance of a spin-$\frac{3}{2}$ system subjected to a periodically-driven magnetic field $\vec{B}(t)$ with angular frequency $\Omega$: 
\begin{equation}
    \hat{H}(t)=(\vec{B}(t)\cdot\hat{\vec{S}})^2,~~~\vec{B}(t)=B(\sin\theta\cos\Omega t,\sin\theta\sin\Omega t,\cos\theta), 
\end{equation}
to demonstrate the quasi-eigenenergy spectroscopy and further estimate the topological holonomy for such dynamics\supercite{zee_non-abelian_1988,Neef2023,Chen2025}. 
The magnitude of the magnetic field is set as $B=2$ in our experiment. We first consider the case of applying a static magnetic field, for which the Hamiltonian yields four eigenstates grouped into two doubly degenerate manifolds with eigenenergies $B^2/4$ and $9B^2/4$, indicated by the two black dotted lines in Figs.~\ref{fig:general}h,\ref{fig:general}i. 
For these two degenerate eigen-spaces, adiabatic transport along a closed loop in the parameter space would formulate a topological holonomy, characterized by the Wilczek-Zee (WZ) phase factor acting on the initial eigenstates \supercite{wilczek_appearance_1984}: 
\begin{equation}
\Phi_{\text{WZ},ab}=\mathcal{P}\text{exp}[i\oint \vec{A}_{\mu,ab}\text{d}x^\mu],
\end{equation}
where $\mathcal{P}$ is path-ordering operator, and $|\phi_a(x)\rangle$ is the eigenstate in a certain degenerate eigen-space and $\vec{A}_{\mu,ab} = i\langle\phi_a(x)|\partial_\mu|\phi_b(x)\rangle$ is the gauge field revealing the geometric structure of the parameter space ($\vec{B}$ in this case). 
Generally, $\Phi_{\text{WZ},ab}$ forms a two-dimensional unitary transformation in the doubly degenerate eigen-space, corresponding to an \text{SU}(2) holonomy. 

To approximate adiabatic evolution and accurately reveal topological phenomena, we choose the driving frequency $\Omega $ as $0.5$. The initial state $\ket{\psi}_T$ is prepared as $e^{-i\theta\hat{S}_y}(0,1,0,0)^T$ for the lower degenerate eigenspace and $e^{-i\theta\hat{S}_y}(0,0,0,1)^T$ for the upper eigenspace, where $\hat{S}_y$ is the four-dimensional representation of the $y$-component spin operator  $\hat{S}$. 
We select the Gaussian window width $\tau$ of 15 to ensure sufficient spectral resolution, with 150 discrete time steps for large spectral range. On the quantum chip, the time-ordered time-evolution operator $\hat{U}(t)=\mathcal{T}e^{-i\int_0^t(\vec{B}(t')\cdot\hat{S})^2\text{d}t'}$ is implemented via the generalised Hadamard test circuits in Fig.~\ref{fig:framework}c. Using UQCS, we measured the quantum auto-correlation function $\mathcal{C}_{\hat{I},\psi}(t)$  and Fourier-transformed eigen-spectra $\widetilde{\mathcal{C}}_{\hat{I},\psi}(\omega)$  for two cases: $\theta=\pi/2$ (Fig.~\ref{fig:general}h) and $\theta=\pi/4$ (Fig.~\ref{fig:general}i), where $\theta$ is  the zenith angle of the magnetic field on the sphere.  
Experimental results reveal the comprehensive Floquet spectrum and topological signatures in this periodically driven systems. 
First, the number of quasi-energy states (represented by blue Gaussian peaks) exceeds the Hamiltonian's dimension, which is a characteristic feature of periodically driven systems as predicted by Floquet theory~\supercite{rudner_band_2020}. Figures~\ref{fig:general}h and \ref{fig:general}i show the Floquet-Bloch band structures (colored bands), where the $\Omega$-periodic replica bands are fully resolved. Our results show excellent agreement with theoretical Floquet spectra. 
Second, the quasi-energy spectra show significant differences between Fig.~\ref{fig:general}h and Fig.~\ref{fig:general}i, particularly pronounced in the first Floquet-Brillouin zone (red-shaded bands), which contains the eigenenergies of the static system. That corresponds to different topological holonomy,
When the gauge field accumulates a scalar phase $\gamma$  during one driving period for states in the degenerate subspace, we observe the expected energy shift  $\Delta E=\gamma\Omega/(2\pi)$ relative to the static system. In  Fig.~\ref{fig:general}h, evolution  along the equator (e.g, $\theta=\pi/2$) creates single shifted Gaussian peaks in both lower and upper level bands, signaling the reduction of SU(2) holonomy to the U(1) Berry phases for each eigenstates \supercite{victor_quantal_1984}. 
This originates from diagonal WZ phase factors, with measured energy shifts of 0.228 for both levels correspond to a topological phase of $0.912\pi$, approaching the predicted $\pi$-phase from adiabatic evolution. 
Notably, evolution along the polar  circle at $\theta=\pi/4$ as shown in  Fig.~\ref{fig:general}i reveals distinct behaviour: 
The double-peak structure  emerges in the  lower-level first band, implying genuine SU(2) holonomy. %
From this double-peak splitting energy $\Delta E_\text{split}$, we estimate the gauge-invariant Wilson loop for the SU(2) holonomy  as $\mathcal{W}=Tr[\Phi_{WZ}]=2\cos(\pi\Delta E_\text{split}/\Omega)=-0.499$, close to the theoretical prediction of $-0.504$. But for the upper level, degenerate eigenstates still acquire U(1) Berry phase as no emergent sub-band energy splitting.
The slight deviation of estimated topological values stems from non-adiabatic coupling between the lower and upper levels. 
Thus, UQCS enables precise computation of Floquet spectra, providing a powerful tool to investigate exotic topological phases and non-equilibrium phenomena in complex quantum systems.



\vspace{2mm}
\noindent
\textbf{Algorithm benchmarking.}
We further benchmark the unique capabilities and performance of UQCS against existing quantum algorithms for spectroscopic estimation of various quantum systems. The UQCS is compared with the iterative quantum phase estimation  algorithm (IQPE) \supercite{PhysRevA.76.030306} and standard QPE,  as well as quantum eigenvalue transformation of unitary matrices (QETU)\supercite{dong2022ground}. Numerical results are reported in Fig.~\ref{fig:benchmark}, and a comprehensive summary   is shown in Table  \ref{tab:benchmark}.

As a test case for Hermitian time-independent quantum system, we adopt the spin chain Hamiltonian  as same as Fig.~\ref{fig:hermitian}f with 8 sites. 
We account for query errors ($\epsilon_q$) arising from imperfections in the execution of single-step controlled unitary operations $e^{i\hat{H}\delta t}$ ($\delta t$ is set as  $0.4$), which are fundamental to eigenenergy estimation in those algorithms. 
Considering experimental errors on the quantum chip, we induce the query error by adding circularly-symmetric complex Gaussian noise to each element of the query unitary: $U_{q,ij}+\epsilon_q*\mathcal{CN}(0,1)$. 
As shown in Fig. \ref{fig:benchmark}a, the error of IQPE grows significantly faster than that of UQCS due to its deeper query depth requirement. 
QETU, based on the fuzzy bisection method \supercite{dong2022ground}, remains robust under small $\epsilon_q$, while it breaks down  when $\epsilon_q$ exceeds the fuzzy interval width of the approximate shifted sign function. 
In contrast, UQCS maintains an estimation error within [$10^{-3}$, $10^{-2}$] thanks to its shallow query depth and the error mitigation technique (singular spectrum renormalisation).  
Figure~\ref{fig:benchmark}b presents the asymptotic scaling of query depth with respect to the target estimation error  ($\epsilon$)  in the absence of query errors. 
The query depth of UQCS scales efficiently  as  $\mathcal{O}(\Delta^{-1}\log(\epsilon^{-1}\zeta^{-1}\Delta))$ (see Methods), where $\zeta=|\bra{\phi_k}\psi\rangle_T|$ denotes the overlap between the target eigenstate $\ket{\phi_k}$ and the trial state $\ket{\psi}_T$   and $\Delta=\Delta E_{min}/R(\hat{H})$ is the rescaled energy gap. Compared to existing algorithms, UQCS achieves optimal query depth.

We numerically  benchmark  the unique capabilities of UQCS for non-Hermitian systems modelled in  Fig.~\ref{fig:general}a. 
We choose the non-collapse eigenstate as the target eigenstate to estimate the real part of eigenenergy through the PT phase transition as shown in Fig.~\ref{fig:general}g. 
Results are shown in Fig.~\ref{fig:benchmark}c. 
While UQCS achieves accurate eigenenergy estimation with minimal error, 
both of IQPE and QETU exbibit  varying degrees of invalidity under the same conditions. 
For QETU, it requries  query to the time-evolution operator $e^{-i\hat{H}_{NH}\delta t}$ as well as its transpose conjugate $e^{i\hat{H}_{NH}^\dagger\delta t}$. Because  the physical eigenstate $\ket{r_k}$ of non-Hermitian Hamiltonian is not an eigenstate of its transpose conjugate $\hat{H}_{NH}^\dagger$, the QETU circuit fails to preserve the dual orthogonal eigenspace. 
This fundamental limitation renders QETU invalid throughout the entire parameter regime for general non-Hermitian eigen-problems. 
The algorithm's relatively small estimation error observed at small $g$ values arises solely because the non-Hermitian Hamiltonian approaches a Hermitian system in this regime.
Moreover, IQPE does not need query to the transpose conjugate of the time-evolution operator, so it can estimate the real eigenenergy in the PT-exact phase. However, two fundamental limitations arise: (1) the presence of additional polynomial terms in the time-evolution operator invalidates IQPE at the EP, and (2) in the PT-broken phase, the imaginary components of eigenenergies induce ancilla qubit amplitude damping,  concealing the phase information.

We also compare various algorithms for the time-varying Floquet quantum system modelled in Figs.~\ref{fig:general}h,i.  
As an example, we estimate Floquet quasi-eigenenergies for the lower degenerate eigenspace with $\theta=\pi/2$, choosing the initial state $\ket{\psi}_T=e^{-i\hat{S}_y\pi/2}(0,1,0,0)^T$. 
Both of IQPE and QETU require a fixed time-evolution operator for their query operations, fundamentally limiting their ability to completely characterise periodically driven systems. 
We examine two evolution times for the query unitaries: the driving period ($\delta t=2\pi/\Omega$), and the discrete time step used in our UQCS experiment ($\delta t= 0.6$). Crucially, IQPE and QETU can only estimate eigenvalues of an effective Hamiltonian  $\hat{H}_{eff}$ defined as $e^{-i\hat{H}_{eff}\delta t}=\mathcal{T}e^{-i\int_0^{\delta t}(\boldsymbol{B}(t')\cdot\hat{S})^2\dd t}$. When  $\delta t$  matches the driving period, $\hat{H}_{eff}$  correctly captures quasi-eigenenergies within a single Floquet-Bloch band \supercite{rudner_band_2020}. However, for other  $\delta t$ values, they yield completely incorrect quasi-eigenenergies. This is confirmed in Fig.~\ref{fig:benchmark}d: only UQCS agrees with theoretical predictions, while IQPE and QETU either return erroneous quasi-eigenenergies or are restricted to estimating a single eigenvalue. 
Additionally, the capability of VQE to solve non-Hermitian eigenproblems across PT symmetric transitions remains uncertain without knowledge of appropriate ansatz. VQE is generally unsuitable for Floquet quantum systems because the effective Hamiltonian describing the full-period time evolution typically lacks a closed form.

\subsection*{Conclusion}
We have demonstrated   a universal quantum computational spectroscopy framework that can efficiently compute spectral properties through quantum auto-correlation function measurements, and have experimentally validated the framework on a programmable silicon-photonic quantum processing chip. 
By leveraging phase orthogonality in quantum dynamics, the framework resolves eigenstates gapped at the real part of eigenenergies, enables eigenstate tomography, detects exceptional points through PT phase transitions, and estimates topological holonomy from Floquet spectra. The multi-functions of UQCS are all experimentally validated on the quantum photonic chip capable of implementing arbitrarily controlled unitaries with high fidelity. 
{Numerical simulations further validate the framework’s unique capability to model general quantum dynamics while maintaining spectral resolution under realistic noise conditions}. Future extensions could integrate classical shadow techniques \supercite{chan_algorithmic_2024} to broaden applicability, employ Laplace transforms \supercite{zylberman2024fastlaplacetransformsquantum} to resolve imaginary part of eigenenergies in non-Hermitian systems, and incorporate wavelet analysis \supercite{Bagherimehrab_2024} for multiscale and time-frequency analysis of non-equilibrium  dynamics. 
Our approach opens avenues to study quantum many-body systems, open quantum systems and non-equilibrium systems beyond the reach of traditional spectroscopy. The synergy of quantum control and spectral analysis presented here could also inspire new paradigms for quantum simulation.

\section*{Methods}
\vspace{-3mm}
\noindent \small{\textbf{Fourier expansion of general quantum dynamics.}
The analysis of the UQCS for general quantum systems is based on the universal Fourier series expansion of the time-evolved initial state: $\hat{U}(t)\ket{\psi}=\sum_n c_n e^{-iE_nt}\ket{u_n}$ with quasi-eigenenergies $E_n$ and quasi-eigenstates $u_n$. We here show the physical meanings of these quantities. The quantum dynamics is generally governed by the Schr\"{o}dinger equation with  Hamiltonian $\hat{H}(t)$ (not necessarily Herimtian):
\begin{equation}
    i\frac{\partial}{\partial t}\ket{\psi(t)}=\hat{H}(t)\ket{\psi(t)}
\end{equation}
which can be Fourier transformed:
\begin{equation}
    \int \text{d}t e^{i\omega t}i\frac{\partial}{\partial t}\ket{\psi(t)}=\int \text{d}t e^{i\omega t}\hat{H}(t)\ket{\psi(t)}
\end{equation}
suppose that the wave function vanishes at infinity, the left hand side can be integrated by parts:
\begin{equation}
    -\int \text{d}t (i\omega)e^{i\omega t}i\ket{\psi(t)}=\int \text{d}t e^{i\omega t}\hat{H}(t)\ket{\psi(t)}
\end{equation}
which can be finally simplified as:
\begin{equation}\label{eqn:M11}
    \omega\ket{u(\omega)}=\int\frac{\text{d}\omega'}{2\pi}\widetilde{H}(\omega-\omega')\ket{u(\omega')}
\end{equation}
where $\ket{u(\omega)}=\int\text{d}te^{i\omega t}\ket{\psi(t)}$ is the Fourier transformed state vector for the time-evolved state, and $\widetilde{H}(\omega)=\int\text{d}te^{i\omega t}\hat{H}(t)$ is the Fourier components of the time-dependent Hamiltonian. 

For most concerned quantum systems, the Fourier components of the Hamiltonian and its generated spectrum are discretised. As such, we consider discretised Fourier decompositions: $\hat{H}(t)=\sum_{m}e^{-i\omega_mt}\widetilde{H}^{(m)}$ ($m\in\mathbb{Z}$) and $\ket{\psi(t)}=\sum_ne^{-iE_nt}\ket{u(E_n)}$ ($m\in\mathbb{Z}$). Note that $\ket{u(E_n)}$ is just the Fourier coefficients vector of $\ket{\psi(t)}$ and not necessarily orthonormal to each other. Then we can obtain the following continuous Fourier transformation: $\widetilde{H}(\omega)=\sum_m 2\pi\delta(\omega-\omega_m)\widetilde{H}^{(m)}$ and $\ket{u(\omega)}=\sum_n2\pi\delta(\omega-E_n)\ket{u(E_n)}$. Substituting this into Eq.\ref{eqn:M11}, we can derive:
\begin{equation}
    \omega\sum_n2\pi\delta(\omega-E_n)\ket{u(E_n)}=\sum_{m,n'}2\pi\delta(\omega-E_{n'}-\omega_m)\widetilde{H}^{(m)}\ket{u(E_{n'})}
\end{equation}


For the gapped real $E_n$, we can integrate the above equation through the interval $(E_n-\epsilon,E_n+\epsilon)$ with the infinitesimal $\epsilon$, which generates:
\begin{equation}\label{eqn:10}
    E_n\ket{u(E_n)} = \widetilde{H}^{(0)}\ket{u(E_n)}+\sum_{m\neq0}\widetilde{H}^{(m)}\ket{u(E_n-\omega_m)}
\end{equation}
Additionally, for any $E_{n'}=E_n-\sum_{l}k_{l}\omega_{l}$ with positive integers $k_l$, which is the modulated energy associated with $E_n$, we can derive that:
\begin{equation}
    E_n\ket{u(E_{n'})} = (\widetilde{H}^{(0)}+\sum_{l}k_{l}\omega_{l})\ket{u(E_{n'})}+\sum_{m\neq0}\widetilde{H}^{(m)}\ket{u(E_{n'}-\omega_m)}
\end{equation}
Stacking up all equations associated with $E_n$ (for any $E_{n'}$), we arrive at an eigen-problem in the extended Hilbert space:
\begin{equation}\label{eqn:12}
  \mathcal{H}\mathcal{U}_n=E_n \mathcal{U}_n 
\end{equation}
\begin{equation}
    \mathcal{H}= \begin{pmatrix}
    \ddots &  &   \\
     & \hat{H}^{(0)}+\sum_l k_l\omega_l-\omega_p & \widetilde{H}^{(p)} & \widetilde{H}^{(p')} &   \\
     & \hat{H}^{(-p)} & \hat{H}^{(0)}+\sum_l k_l\omega_l & \hat{H}^{(p)} & \\
     & \hat{H}^{(-p')} & \hat{H}^{(-p)} &  \hat{H}^{(0)}+\sum_l k_l\omega_l+\omega_p & \\
     & & & & \ddots
    \end{pmatrix}
\end{equation}
\begin{equation}
    \mathcal{U}_n = \begin{pmatrix}
      \vdots\\\ket{u(E_{n'}+\omega_p)} \\ \ket{u(E_{n'})} \\ \ket{u(E_{n'}-\omega_p)} \\ \vdots
    \end{pmatrix}
\end{equation}
where $p$ is a integer, $\widetilde{H}^{(-p)}$ denotes the Fourier component of Hamiltonians corresponding to $e^{i\omega_pt}$ and $\widetilde{H}^{(p')}$ ($\widetilde{H}^{(-p')}$) represents the component with the frequency $\omega_{p'}=2\omega_p$ ($\omega_{p'}=-2\omega_p$).
We only show the interactions between three Fourier components, but in general the eigen-problem can be infinite-dimensional and we usually truncate those components in high-frequency for good enough approximation.

The $E_n$ from solving the Eq.~(\ref{eqn:12}) together with its modulated energy $E_{n'}=E_n-\sum_lk_l\omega_l$ constructs the full spectrum of general quantum systems, all of which we call quasi-eigenenergies denoted in the same form $E_n$ (including all $E_{n'}$). Associated with each quasi-eigenenergy $E_n$, the solved $\ket{u_n}=\ket{u(E_n)}$ is named as quasi-eigenstate and obviously not necessarily orthonormal to each other. Hence, any time-evolved state can be expanded by these Fourier components $\hat{U}(t)\ket{\psi}=\sum_n c_n e^{-iE_nt}\ket{u_n}$, and for the unit-norm condition of quantum states, we induce additional coefficients $c_n$ to normalise the time-evolved state. 

For the time-independent Hamiltonian, the Eq.~(\ref{eqn:10}) degrades into the traditional eigen-problem in the stationary Schr\"{o}dinger equation
\begin{equation}
E_n\ket{u(E_n)}=\widetilde{H}^{(0)}\ket{u(E_n)}=\hat{H}\ket{u(E_n)}
\end{equation}
with eigenerengy $E_n$ and eigenstate $\ket{u(E_n)}$. For the periodically-driven Hamiltonian with driven frequency $\Omega$, $\omega_m=m\Omega$ ($m \in \mathbb{Z}$) and $E_n$ is associated with $E_{n'}=E_n-m\Omega$, which forms the repetitive Floquet-Bloch bands separated by $\Omega$. 
The distinct eigenvalues of Eq.~(\ref{eqn:12}) generate the sub-band spectrum in a single band, as shown in Fig.~\ref{fig:general}i.
}

\vspace{2mm}
\noindent\small{\textbf{Query depth.} 
{UQCS efficiently scales as $\mathcal{O}[(\sqrt{2\ln(1/\epsilon_1)}R(\hat{H})/\Delta E_{min})^2/\epsilon_2^2]$. The quantity $\sqrt{2\ln(1/\epsilon_1)}R(\hat{H})/\Delta E_{min}$ represents the required points in the discrete-time Fourier transformation for the required integral truncation error $\epsilon_1$. 
Since we consider the query unitary as the single-step time-evolution unitary, this quantity is propotional to the query depth. Then we need translate the Fourier integral truncation error to the energy estimation error. For ideal case, the Gaussian peak in the spectrum is $f(\omega)=\zeta^2\exp(-\frac{1}{2}\tau^2(\omega-E_n)^2)$. Then we can obtain the dependence of energy error $\delta \omega$ on the function error $\delta f$ (i.e. Fourier integral error) near the peak: $|\delta\omega|^2\approx2|\delta f|/(\zeta^2\tau^2)$. The quantity can be translated to $\sqrt{2\ln(2/(\epsilon^2\zeta^2\tau^2))}R(\hat{H})/\Delta E_{min}$. With relation of the rescaled energy gap $\Delta=\Delta E_{min}/R(\hat{H})$ and $\tau\sim1/\Delta$, we can obtain the query depth of UQCS scales as $\mathcal{O}(\Delta^{-1}\sqrt{\log(\epsilon^{-1}\zeta^{-1}\Delta)})$.}

\vspace{2mm}
\noindent\small{\textbf{Device fabrication.} Integrated silicon-photonic quantum chips implementing the generalized Hadamard test were fabricated using standard  CMOS processes. The waveguide-based quantum circuits (Fig.~\ref{fig:chip}a) were patterned on an 8-inch silicon-on-insulator (SOI) wafer through 248 nm deep ultraviolet (DUV) photolithography and inductively coupled plasma (ICP) etching. Following waveguide fabrication, a 1 $\mu$m-thick silicon dioxide layer was deposited via plasma-enhanced chemical vapor deposition (PECVD). A 50-nm-thick titanium nitride  layer was then deposited atop the waveguides to serve as thermo-optic phase shifters. The silicon waveguides feature a cross-section of $450~nm\times220~nm$. Photon-pair sources with 1.5 cm-long waveguide sections were designed to generate sufficiently bright entangled photons. Balanced beamsplitters were implemented using multimode interferometers  measuring $27~\mu m\times2.8~\mu m$  The chip is  is controlled by a multi-channel digital-to-analog converter  driver with a maximum output voltage of 12 V.}

\vspace{2mm}

\noindent \small{\textbf{Experimental setup.} 
In our experiment, a tunable continuous-wave laser centered at 1550.12 nm pumped the waveguide photon sources after amplification to 200 mW using an erbium-doped fiber amplifier (EDFA). Photon pairs were generated via spontaneous four-wave mixing (SFWM) in integrated sources and subsequently separated by on-chip asymmetric Mach-Zehnder interferometers (AMZIs), with signal and idler photons at 1545.32 nm and 1554.94 nm, respectively. The photons were coupled off-chip via grating couplers and detected by fiber-coupled superconducting nanowire single-photon detectors (SNSPDs, average efficiency: 85\%). Coincidence counts were recorded using a multichannel time-interval analyzer. Photon rates varied with the prepared quantum states and measurement bases. For each generalized Hadamard test expectation value estimation, we used 5 second integration time in projective measurements to achieve approximately 1000 two-photon coincidences, ensuring sufficiently small shot noise. }

\section*{Data Availability}
\normalsize{The data that support the findings of this study are available from the corresponding author upon request.}

\section*{Code Availability}
\normalsize{The numerical simulation of noisy quantum circuits is performed by MindSpore Quantum\supercite{xu2024mindspore}, the codes of which are available on GitHub \href{https://github.com/Celesoar/MindQuantum-for-UQCS.git}{https://github.com/Celesoar/MindQuantum-for-UQCS.git}.}

\printbibliography

\section*{Acknowledgements} 
{We acknowledge support from the National Natural Science Foundation of China (nos.~123B2065, 12325410, 62235001, 11834010, 12361161602), NSAF (no.~U2330201), the Innovation Program for Quantum Science and Technology (nos.~2021ZD0301500,  2023ZD0300200), the Beijing Natural Science Foundation (Z220008), the High-performance Computing Platform of Peking University, and the MindSpore Quantum framework for numerical noisy quantum simulations. 
J.S. would like to thank support from the Innovate UK (project no.~10075020) and support through Schmidt Sciences, LLC. 
}

\section*{Authors contributions} 
C.Z. and J.S. conceived the theoretical framework. C.Z., J.H. and J.W. designed the quantum photonic devices. C.Z., J.H., J.M., H.B. and S.Z. implemented the experiment. C.Z. provided the simulations and performed the noise analysis. 
Q.G., V.V., X.Y. and J.W. supervised the project. C.Z., J.S., X.Y. and J.W. wrote the manuscript with input from all the authors. All the authors discussed the results and contributed to the manuscript.

\section*{Competing interests} 
The authors declare no competing interests.

\end{document}